\documentclass[aps, pra, superscriptaddress, twocolumn]{revtex4-1}

\usepackage{amsmath, amssymb, amsthm, mathtools, mathrsfs}
\usepackage{physics}
\usepackage{color}

\usepackage{graphicx}
\usepackage{lipsum}
\usepackage{numprint}
\usepackage{enumitem}

\usepackage[per-mode=symbol]{siunitx}
\DeclareSIUnit\Cu{Cu}
\usepackage{comment}

\usepackage{hyperref}

\newcommand{\lsco}{La${}_{2-x}$Sr$_{x}$CuO${}_4$}
\newcommand{\lbco}{La${}_{2-x}$Ba$_{x}$CuO${}_4$}

\usepackage{ulem}

\begin{document}
	\title{Light-induced nonequilibrium response of the superconducting cuprate \lsco{}}
	\author{Hiroaki~Niwa}
	\affiliation{Department of Physics, The University of Tokyo, Hongo, Tokyo 113-0033, Japan}
	
	\author{Naotaka~Yoshikawa}
	\affiliation{Department of Physics, The University of Tokyo, Hongo, Tokyo 113-0033, Japan}
	
	\author{Kaito~Tomari}
	\affiliation{Department of Physics, The University of Tokyo, Hongo, Tokyo 113-0033, Japan}

	\author{Ryusuke~Matsunaga}
	\affiliation{Department of Physics, The University of Tokyo, Hongo, Tokyo 113-0033, Japan}
	\affiliation{Laser and Synchrotron Research Center, The Institute for Solid State Physics, The University of Tokyo, 5-1-5 Kashiwanoha, Kashiwa, Chiba, 277-8581, Japan}
	
	\author{Dongjoon~Song}
	\affiliation{National Institute of Advanced Industrial Science and Technology, Tsukuba 305-8568, Japan}
	
	\author{Hiroshi~Eisaki}
	\affiliation{National Institute of Advanced Industrial Science and Technology, Tsukuba 305-8568, Japan}
	
	\author{Ryo~Shimano}
	\affiliation{Department of Physics, The University of Tokyo, Hongo, Tokyo 113-0033, Japan}
	\affiliation{Cryogenic Research Center, The University of Tokyo, Yayoi, Tokyo 113-0032, Japan}
	\date{\today}

	\begin{abstract}
	We report the dynamics of the cuprate superconductor \lsco{}~$(x = 0.14)$  after intense photoexcitation utilizing near-infrared (\SI{800}{\nano\meter}) optical pump-terahertz probe spectroscopy. In the superconducting state at \SI{5}{\kelvin}, we observed a redshift of the Josephson plasma resonance that sustains for hundreds of picoseconds after the photoexcitation, indicating the destruction of the $c$-axis superconducting coherence. We show that the metastable spectral features can be described by the photoinduced surface heating of the sample. We also demonstrate that the conventional analysis used to extract the spectra of the photoexcited surface region can give rise to artifacts in the nonequilibrium response.
	\end{abstract}

	\maketitle
	\section{Introduction}
	In unconventional high-temperature superconductors, it has been revealed that various electronic states emerge depending on temperature and chemical doping, such as antiferromagnetic insulator, spin density wave, stripe order, charge order, pair density wave, nematic order, and pseudogap~\cite{Keimer2015}. It is now commonly recognized that the elucidation of the interplay of those multiple orders as well as the unveiling of pairing glue is crucial to understand the emergence of superconductivity. In these aspects, the study of real-time dynamics of superconductivity by ultrafast spectroscopy technique has been playing important roles~\cite{Giannetti2016,DalConte2012,Sentef2013}. For instance, ultrafast pump-probe experiments have shown their ability to elucidate bosonic fluctuations to which fermionic quasiparticles couple. Time-resolved observation of collective modes enables the direct access to the order parameter dynamics in nonequilibrium~\cite{Matsunaga2013,Matsunaga2014,Katsumi2018}, and for cuprate superconductors it has been used to reveal the coupling between the superconductivity and the charge density wave order~\cite{Dakovski2015,Hinton2016}. 
	
	Searching for a new state of matter induced by the photoexcitation is also a fascinating subject in this respect, as it provides deeper insight into the competing orders or hidden states, with revealing new functionalities of correlated materials. Light-induced superconductivity is a highly intriguing example, where the strong photoexcitation leads to a transient emergence of superconductivity-like response above the critical temperature $T_\mathrm{c}$~\cite{Fausti2011,Kaiser2014a,Hu2014,Nicoletti2014,Casandruc2015,Nicoletti2018}. For example in \lbco , transient reappearance of Josephson plasma resonance (JPR), one of the characteristic fingerprint of superconductivity, has been observed in the $c$-axis terahertz (THz) response above $T_\mathrm{c}$ right after intense near-infrared photoexcitation~\cite{Nicoletti2014,Casandruc2015}. It has been suggested that the photoexcitation suppresses the competing stripe charge order~\cite{Nicoletti2014,Casandruc2015,Khanna2016}, although the microscopic origin of this light-induced phenomenon remains to be resolved. The photoexcited state of \lbco~below $T_\mathrm{c}$ is even more unclear. Though a transient enhancement of JPR frequency was primarily reported~\cite{Nicoletti2014}, recent experimental results have been suggesting an existence of photoinduced metastable phase, where a new JPR mode and significant increase of spectral weight in optical conductivity are observed~\cite{Zhang2018,Cremin2019}. Hence, the role of intense photoexcitation and its relation with the preexisting competing orders remain to be clarified.
	
	Here, we investigated the photoexcited dynamics in the nearly optimal-doped \lsco{}~$(x = 0.14)$,  one of the archetypal cuprate superconductors, by utilizing near-infrared optical-pump and THz-probe spectroscopy. In \lsco, the effect of charge stripe order has been reported to be less pronounced compared to \lbco{}~\cite{Vojta2009,Wu2012}. In addition, previous ultrafast pump-probe studies of \lsco~have presented a destruction of superconductivity after the photoexcitation~\cite{Kusar2005, Kusar2008, Beyer2011}. Thus, \lsco~provides an interesting platform to expose how the significance of stripe charge order affects the photoexcited state below $T_\mathrm{c}$. With increasing excitation density, the JPR shifts to the low energy side which sustains for several hundreds of picoseconds, indicating a quasiequilibrium state with suppressed bulk superconducting coherence. We show that this quasiequilibrium state is well-explained by the thermalization due to the photoexcitation. At the same time, we argue the breakdown of conventional analysis used in optical pump-THz probe experiments, which is prone to produce nonexistent features in the terahertz-range spectra.
	
	\section{Experiments}
	\subsection{Methods}
	We used a bulk \lsco{} single crystal with the doping level of $x = 0.14$ grown by floating-zone method. The mirror-polished $ac$ surface of the sample (\SI[product-units = power]{7x7x5}{\milli\meter}) was used and masked by a metal plate with a tapered 4-\si{\milli\meter} hole. A gold mirror was also mounted on the sample holder as a reference for the reflectivity measurement by THz time-domain spectroscopy (THz-TDS). In Fig.~\ref{fig:figure_1}(a) we show a schematic of near-infrared optical pump-THz probe spectroscopy. As a light source we used a Ti:Sapphire-based regenerative amplifier with the pulse energy of \SI{4.2}{\milli\joule}, repetition rate of \SI{1}{\kilo\hertz}, pulse duration of \SI{100}{\femto\second}, and center wavelength of \SI{800}{\nano\meter}. The output of the laser was divided into three beams; each for the optical pump, the generation of the THz probe, and the gate pulse for the THz-TDS, respectively. The optical pump beam has a Gaussian profile with $1/e^2$ diameter of \SI{9}{\milli\meter}, which ensures spatial uniformity of excitation density on the probed region. The THz pulse was generated by the optical rectification in a large-aperture ZnTe crystal and linearly polarized along the $c$-axis of \lsco{}. The reflected THz pulse was detected by the electro-optic sampling in a ZnTe crystal with the gate pulse. The waveform of the probe THz pulse is shown in Fig.~\ref{fig:figure_1}(b).

	\subsection{Equilibrium properties}
	\begin{figure}
		\includegraphics[width=\columnwidth]{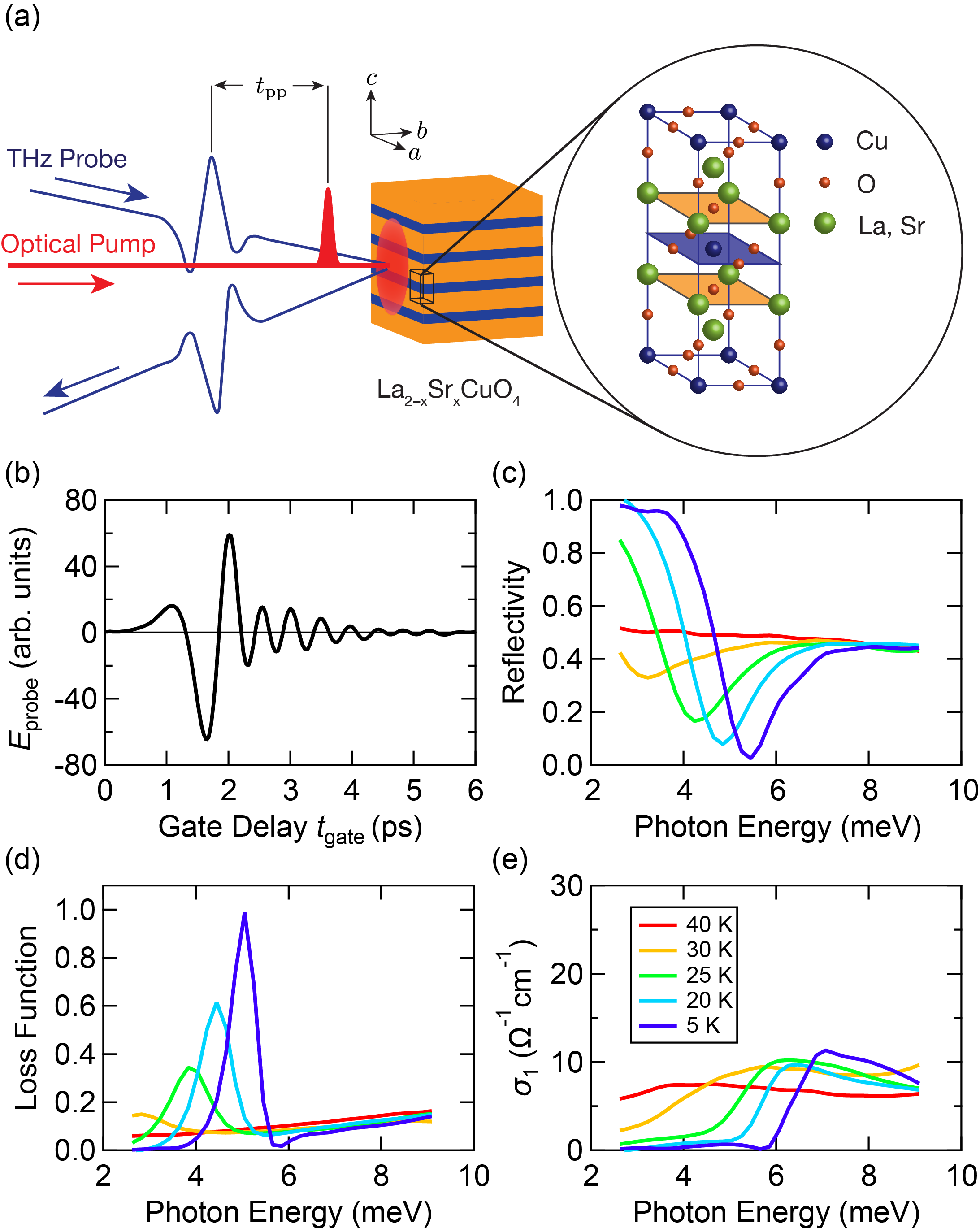}
		\caption{(Color online)~Experimental setup and optical properties in equilibrium. (a)~Schematic of the optical pump-THz probe experiment and the crystal structure of \lsco{}. Both pump and probe pulses are polarized along the $c$-axis direction of the crystal. (b)~Time-domain waveform of the probe THz pulse generated from ZnTe. (c)-(e)~Reflectivity, loss function and real-part optical conductivity of \lsco{} $(x = 0.14)$ from \SI{5}{\kelvin} to \SI{40}{\kelvin}, respectively.}
		\label{fig:figure_1}
	\end{figure}  
	Figures~\ref{fig:figure_1}(c)-(e) show the optical properties of \lsco{}~$(x=0.14)$ in equilibrium. In the reflectivity spectrum (Fig.~\ref{fig:figure_1}(c)), a sharp plasma edge associated with the JPR is discerned below $T_\mathrm{c}$. Concomitantly, the loss function spectrum (Fig.~\ref{fig:figure_1}(d)) as defined by $-\Im(1/\varepsilon(\omega))$ with $\varepsilon(\omega)$ the complex dielectric function exhibits a single peak, corresponding to the longitudinal JPR mode. As temperature decreases the plasma edge shows a blueshift and approaches to \SI{5.5}{\milli\electronvolt} reflecting the development of $c$-axis superconducting coherence, which is in good agreement with the previous studies~\cite{Tamasaku1992,Uchida1996,Basov1995,Dordevic2003}. A notable feature is identified in the $c$-axis real-part optical conductivity $\sigma_1(\omega)=\Re[\sigma(\omega)]$ as shown in Fig.~\ref{fig:figure_1}(e); the low-frequency spectral weight is gradually suppressed below $T_\mathrm{c}$ and exhibits gap-like behavior at the lowest temperature. This spectral feature has also been identified in previous reports on the $c$-axis optical response of \lsco{}~\cite{Tamasaku1992,Basov1995,Dordevic2003}. As the origin of this spectral peak in $\sigma_1(\omega)$, two scenarios have been considered: 1)~the transverse Josephson plasma mode activated by inhomogeneous distribution of inter-layer Josephson coupling constant~\cite{VanderMarel1996,Gruninger2000}, and 2)~the in-plane inhomogeneity of superconductivity associated with the spin-density wave~\cite{Basov1995,Dordevic2003}, while its assignment remains an open issue.
	
	\section{Results of Photoexcited State}
\begin{figure}
		\includegraphics[width=\columnwidth]{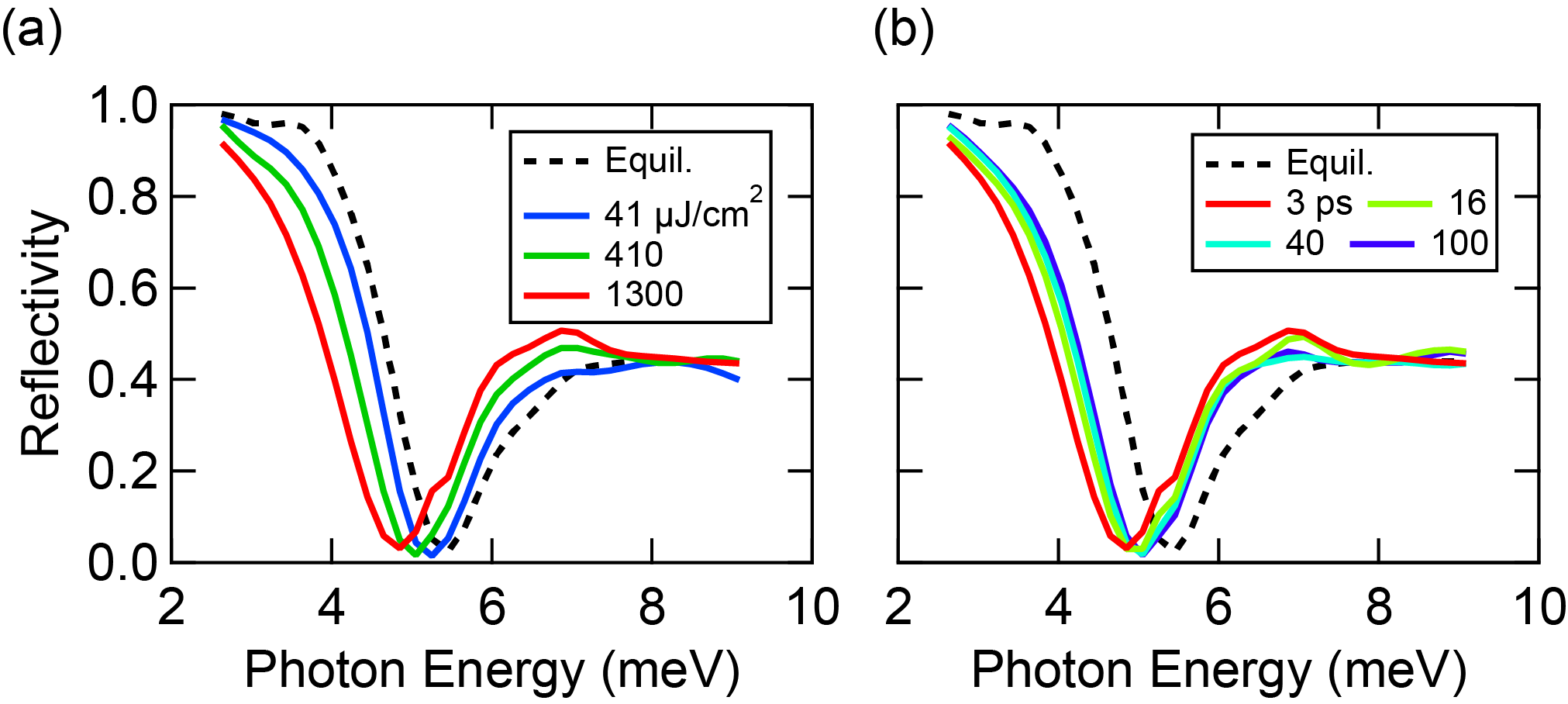}
		\caption{(Color online)~(a)~Pump-fluence dependence of transient reflectivity at $t_\mathrm{pp} = \SI{3}{\pico\second}$. (b)~Time evolution of transient reflectivity at the pump fluence of \SI{1300}{\micro\joule\per\square\centi\meter}.}
		\label{fig:figure_2}
	\end{figure}

	Now we show the result of near-infrared optical pump-THz probe spectroscopy of \lsco{}~$(x = 0.14)$ with the optical pump linearly polarized along the $c$-axis. We also investigated the pump polarization along the $a$-axis and obtained similar results. Figure~\ref{fig:figure_2}(a) shows transient reflectivity measured at \SI{3}{\pico\second} after the optical pump ($t_\mathrm{pp}= \SI{3}{\pico\second}$). 
	As the pump fluence increases, the JPR plasma edge shows a gradual redshift, indicating the reduction of $c$-axis superconducting coherence. Figure \ref{fig:figure_2}(b) shows the time evolution of transient reflectivity at the photoexcitation density of \SI{1300}{\micro\joule\per\square\centi\meter}. The pump-induced redshift of JPR is most significant right after the excitation $(t_\mathrm{pp}=\SI{3}{\pico\second})$, and shows slow recovery toward the equilibrium plasma frequency. The spectral features at $t_\mathrm{pp}=\SI{40}{\pico\second}$ and $t_\mathrm{pp}= \SI{100}{\pico\second}$ are almost identical, suggesting that at $t_\mathrm{pp}= \SI{40}{\pico\second}$ the system has reached quasiequilibrium. A hump structure in reflectivity emerges around \SIrange[range-units = single, range-phrase = --]{6}{7}{\milli\electronvolt} and shows the same relaxation behavior in terms of its timescale.
	
	Here, we compare the nonequilibrium dynamics with previous studies. Ultrafast dynamics of the photoexcited \lsco{} has been studied by transient reflectivity measurements at \SI{800}{\nano\meter}~\cite{Kusar2005,Kusar2008,Beyer2011} and by THz transmittance measurements~\cite{Beyer2011}. In optical pump-probe measurements using a thin-film \lsco{}, the excitation density required to destroy the superconductivity at $t_\mathrm{pp}= \SI{3}{\pico\second}$ has been estimated to be \SI{2.4}{\kelvin\per\Cu}~\cite{Beyer2011}, which corresponds to approximately \SI{30}{\micro\joule\per\square\centi\meter} in our current experimental setup. Therefore, we expect that in our result shown in Fig.~\ref{fig:figure_2}(a) the superconductivity is destroyed in all three pump fluences at $t_\mathrm{pp}= \SI{3}{\pico\second}$. The transient reflectivity in Fig.~\ref{fig:figure_2}, however, shows only the slight redshift of JPR without complete suppression. It can be ascribed to the penetration-depth mismatch of the pump and probe pulses. The penetration depth of the THz probe is on the order of \SI{10}{\micro\meter}, whereas the penetration depth at \SI{800}{\nano\meter} is around \SI{660}{\nano\meter} as obtained from the literature values~\cite{Uchida1996}. Therefore, in optical pump-THz probe measurements  in reflection geometry, what we observe is the mixed response of small contribution from the surface photoexcited region on top of the large contribution from the equilibrium background. The consideration of the penetration-depth mismatch will be discussed in more detail in the later section.
	 
	The intense photoexcitation leads to the heating of the sample, which has been a well-known contribution in pump-probe experiments. The pump energy density required to destroy the superconductivity at $t_\mathrm{pp} = \SI{100}{\pico\second}$ after the photoexcitation, where the electronic system and the lattice system are considered to reach the temperature just above $T_\mathrm{c}$, has been evaluated as \SI{13.7}{\kelvin\per\Cu} from the saturation behavior of the pump-probe signal~\cite{Beyer2011}. This photoexcitation intensity corresponds to approximately \SI{150}{\micro\joule\per\square\centi\meter}, thus for the pump fluence of \SI{410}{\micro\joule\per\square\centi\meter} and above, we expect that photoexcited sample has been heated up above $T_\mathrm{c}$ and the superconductivity is thermally destroyed. 
	 
	\section{Analysis of the photoexcited state}\label{sec:analysis}
	\subsection{Modeling of the quasiequilibrium state}\label{sec:modeling}
\begin{figure}
		\includegraphics[width=\columnwidth]{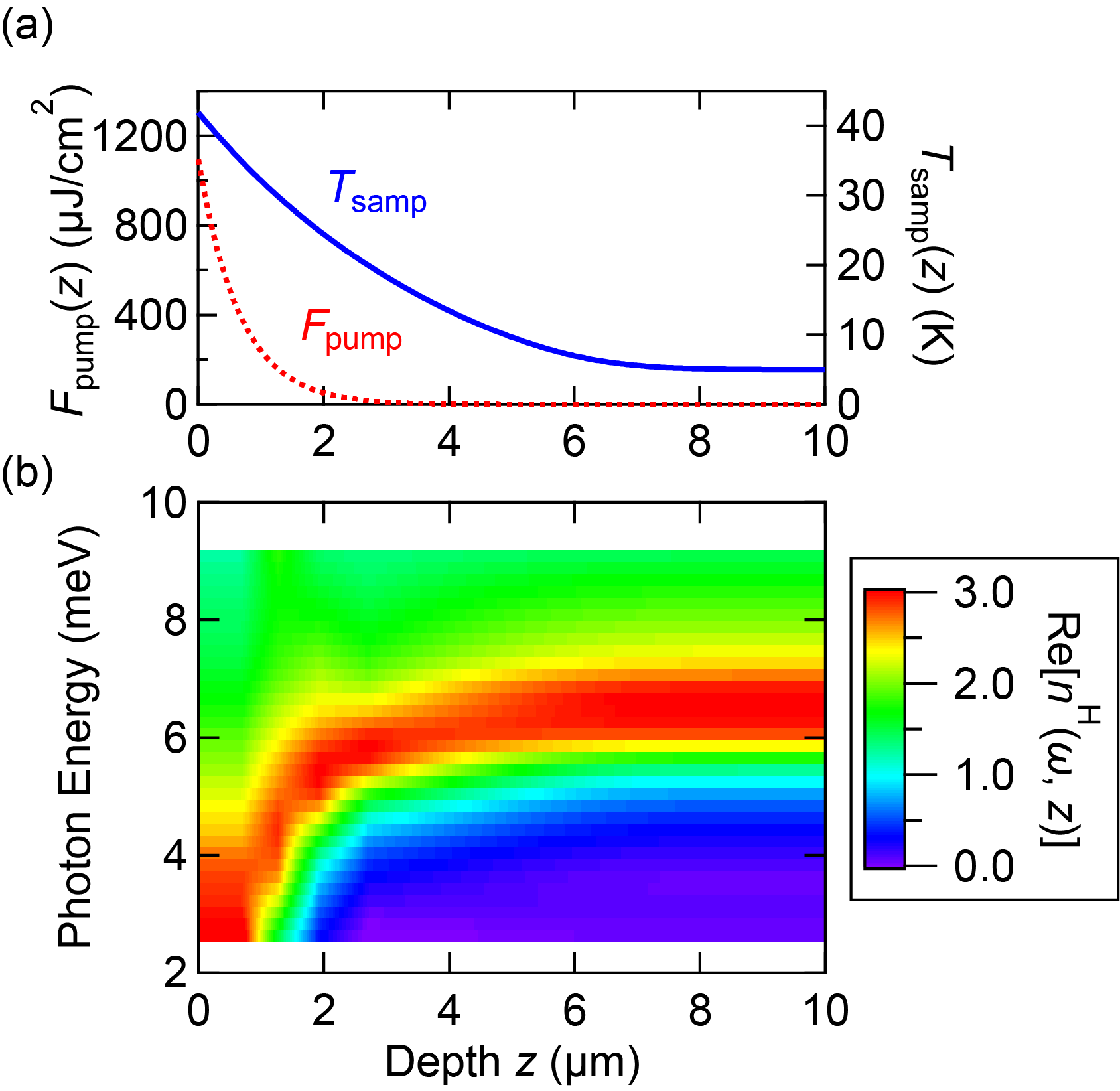}
		\caption{(Color online)~Simulation of the sample heating induced by the optical excitation. (a)~Pump fluence $F_\mathrm{pump}(z)$ at \SI{1300}{\micro\joule\per\square\centi\meter} excitation (red dotted line) and the corresponding temperature $T_\mathrm{samp}(z)$ (blue solid line) at the sample depth $z$. (b)~Constructed spatial distribution of the real-part refractive index $\Re[n^\mathrm{H}(\omega,z)]$.}
		\label{fig:figure_3}
	\end{figure}

	To examine the effect of pump-induced heating in quasiequilibrium state, we consider a following model of spatial distribution of the refractive index $n^\mathrm{H}(\omega, z)$, assuming that all of the injected energy by photoexcitation is converted to the temperature increase in the surface region of the sample. Utilizing the result of specific heat measurement~\cite{Wen2004}, one can convert the pump-energy density to the amount of temperature increase at each sample depth $z$ (See Appendix~\ref{sec:heatsim} for the details of calculation.). In Fig.~\ref{fig:figure_3}(a), we show the fluence of the optical pump $F_\mathrm{pump}(z)$ for \SI{1300}{\micro\joule\per\square\centi\meter} excitation and corresponding quasiequilibrium temperature $T_\mathrm{samp}(z)$ obtained by converting the pump-energy density to the local temperature. Notably, we see that the photoexcitation affects the sample temperature much more deeply into the sample compared to the pump penetration depth; $F_\mathrm{pump}(z)$ shows an exponential decay in the length scale of the pump penetration depth, but $T_\mathrm{samp}(z)$ slowly recovers back to its original equilibrium temperature in \SI{6}{\micro\meter}, mainly due to the small specific heat at low temperature. 

	From the obtained temperature distribution $T_\mathrm{samp}(z)$, it is now possible to construct $n^\mathrm{H}(\omega, z)$ by assigning equilibrium data of the refractive index at the corresponding temperature for each $z$, which is presented in Fig.~\ref{fig:figure_3}(b). The total reponse of the sample with $n^\mathrm{H}(\omega, z)$ when observed by the THz pulse can be obtained as the effective refractive index $n_\mathrm{eff}^\mathrm{H}(\omega)$, by assuming the surface region of the sample as a stack of thin layers and calculating the Fresnel coefficients considering the multiple reflections (See Appendix~\ref{sec:n_eff} for the details of calculation.). In Fig. 4, we compare between the energy reflectivity measured at $t_\mathrm{pp}=\SI{100}{\pico\second}$ and the calculated energy reflectivity from the simulation of $n_\mathrm{eff}^\mathrm{H}(\omega)$ with various pump fluences. The simulation result describes the experimental data in quasiequilibrium state consistently. The growth of the hump structure around  \SIrange[range-units = single, range-phrase = --]{6}{7}{\milli\electronvolt} is also apparent in the simulation. Comparing the simulated spectra with the measurement at $t_\mathrm{pp}=\SI{3}{\pico\second}$ in Fig.~\ref{fig:figure_2}(a), the experimental result shows more pronounced redshift of JPR and the growth of hump structure. At earlier pump-probe delaytime, application of the heating simulation may not be appropriate because the sample is in the middle of the thermalization processes to reach the quasiequilibrium state and the sample temperature cannot be clearly defined. Nevertheless, the spectral features in quasiequilibrium state in photoexcited \lsco{} observed at longer delaytime can be understood by considering the sample heating due to the energy injection by the optical pump.
	
	\begin{figure}
		\includegraphics[width=\columnwidth]{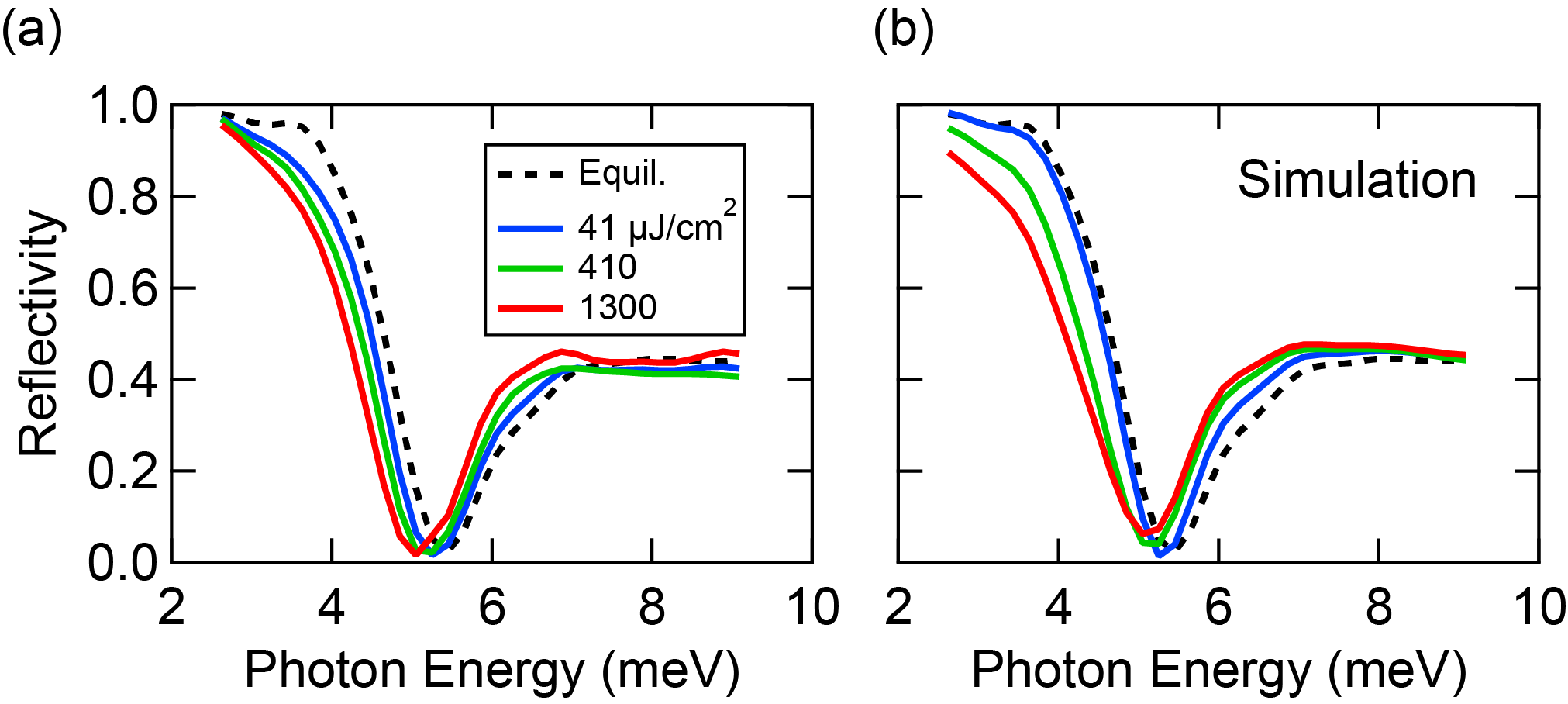}
		\caption{(Color online)~Comparison of the experimental result with the heating simulation. (a)~Pump-fluence dependence of energy reflectivity at $t_\mathrm{pp} = \SI{100}{\pico\second}$ and (b)~corresponding energy reflectivity obtained from the heating simulation. }
		\label{fig:figure_4}
	\end{figure}
	\subsection{Possible artifacts produced in conventional pump-probe analysis}
	To gain further insight toward the spectra in the photoexcited state, we must consider the penetration-depth mismatch between pump and probe in order to obtain the response functions such as the loss function and optical conductivity. In previous studies of optical pump-THz probe experiments in reflection geometry, this discrepancy was taken into account by postulating a function form for the spatial distribution of complex refractive index $n(\omega, z)$. One example is the multilayer analysis~\cite{Kaiser2014a,Zhang2018}, where one assumes that the photoinduced change of the spatial profile of the complex refractive index $n^\mathrm{ML}(\omega, z)$ can be written by the exponential decay with respect to the pump penetration depth $d_\mathrm{pump}$ i.e.,
	\begin{align}
	n^\mathrm{ML}(\omega, z) = n_\mathrm{eq}(\omega) + (n_\mathrm{surf}(\omega) - n_\mathrm{eq}(\omega))e^{-z/d_\mathrm{pump}},
	\end{align}
	where $n_\mathrm{eq}(\omega)$ and $n_\mathrm{surf}(\omega)$ are the equilibrium and photoexcited surface complex refractive index, respectively. One can then calculate the effective refractive index $n^\mathrm{ML}_\mathrm{eff}(\omega)$, which is performed in the same manner as the heating simulation of quasiequilibrium state in Sec.~\ref{sec:modeling}, and obtain the value of $n_\mathrm{surf}(\omega)$ such that $n^\mathrm{ML}_\mathrm{eff}(\omega)$ fits the complex refractive index obtained from experiments. Another method is the single-layer analysis~\cite{Hu2014,Nicoletti2014,Casandruc2015,Cremin2019}, where one assumes that the photoexcitation creates a thin surface nonequilibrium layer with the thickness of $d_\mathrm{pump}$ on top of the equilibrium bulk, and includes all of the photoinduced effects into the surface thin layer i.e.,
	\begin{align}
	n^\mathrm{SL}(\omega, z) = \begin{cases}
	n_\mathrm{surf}(\omega)\ &(z \leq d_\mathrm{pump}),\\
	n_\mathrm{eq}(\omega)\ &(\text{otherwise}).
	\end{cases}
	\end{align}
	As far as the extraction of $n_\mathrm{surf}(\omega)$ is concerned, both procedures are known to produce similar spectral features at the sample surface~\cite{Nicoletti2014,Casandruc2015,Zhang2018}.
	\begin{figure}
		\includegraphics[width=\columnwidth]{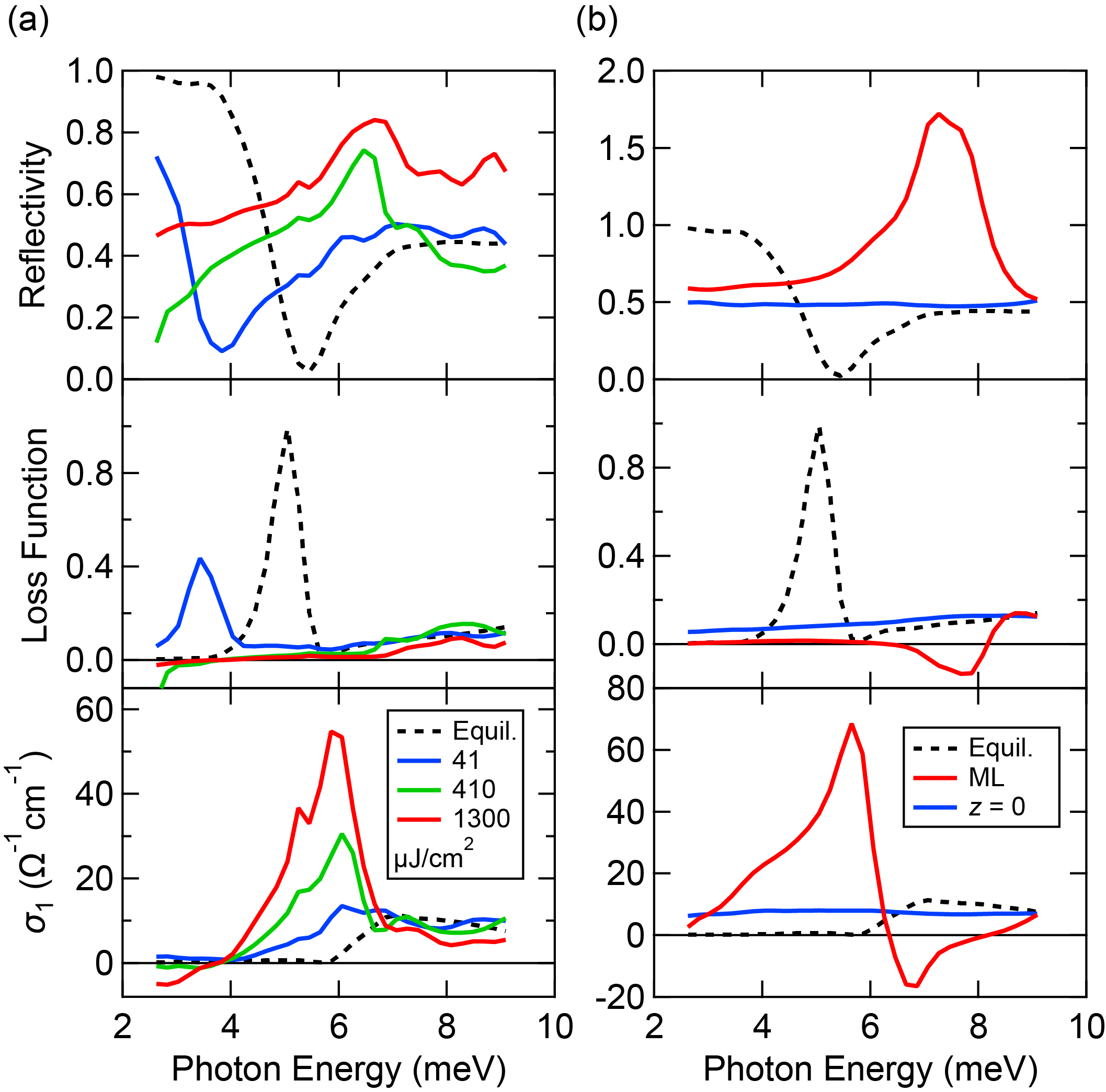}
		\caption{(Color online)~(a)~Pump-fluence dependence of reflectivity, loss function and real-part optical conductivity at $t_\mathrm{pp} = \SI{100}{\pico\second}$ extracted after the multilayer analysis. (b)~Reflectivity, loss function and real-part optical conductivity obtained from the multilayer analysis of $n^\mathrm{H}_\mathrm{eff}(\omega)$  for \SI{1300}{\micro\joule\per\square\centi\meter} excitation shown in Figs. \ref{fig:figure_4}(b) (red), and corresponding spectra calculated from $n^\mathrm{H}(\omega, z = 0)$ of the heating simulation (blue). }
		\label{fig:figure_5}
	\end{figure}
	Here, we applied the former multilayer analysis and reconstructed the transient optical spectra at the photoexcited sample surface based on the obtained $n_\mathrm{surf}(\omega)$.
	
	In Fig.~\ref{fig:figure_5}(a), we present the fluence dependence of the energy reflectivity, loss function and real-part optical conductivity at $t_\mathrm{pp}= \SI{100}{\pico\second}$ after extracting the surface refractive index by applying the multilayer analysis to the experimental result shown in Fig.~\ref{fig:figure_4}(a). The superconducting JPR shows a strong redshift even in the weak excitation regime (\SI{41}{\micro\joule\per\square\centi\meter}), and it is completely suppressed for stronger excitation density. In the real-part conductivity, a sharp peak emerges from the edge of the conductivity gap and evolves into a prominent peak with increasing pump fluence.
	
	However, as we demonstrated in Sec.~\ref{sec:modeling} and Fig.~\ref{fig:figure_4}(b), the nonequilibrium optical spectra are reproduced by taking into account the pump-induced heating effect. Supposing the effective refractive index from the heating simulation $n^\mathrm{H}_\mathrm{eff}(\omega)$ as an experimental result and applying the multilayer analysis, we can now compare the extracted photoexcited surface spectra obtained from $n_\mathrm{surf}(\omega)$ with the spectra at $z=0$ of the heating simulation, which can be calculated using $n^\mathrm{H}(\omega, z = 0)$. We show the comparison of the extracted surface spectra and the simulated data for \SI{1300}{\micro\joule\per\square\centi\meter} excitation in Fig.~\ref{fig:figure_5}(b). The spectra calculated from $n_\mathrm{surf}(\omega)$ of the simulation result also present the similar behavior of increased energy reflectivity and corresponding increase in spectral weight in the real-part optical conductivity. The effect is more exaggerated in these spectra; the reflectivity exceeds one and the loss function becomes negative. On the other hand, the actual surface spectra obtained from $n^\mathrm{H}(\omega, z=0)$ of heating simulation show featureless reflectivity, loss function and optical conductivity. This comparison demonstrates a possibility that the conventional method to extract the surface response functions does not provide the actual spectra at the surface but instead resulting in artificial spectral properties. 
	
	The observation of artifacts presented in Fig.~\ref{fig:figure_5} directly points out the fragility of surface reconstruction methods. As we can clealy see in Fig.~\ref{fig:figure_3}, the spatial distribution of refractive index does not follow the exponential decay. The validity of the function form of $n(\omega, z)$, both in single-layer and multilayer analysis, may break down and lead to artifacts especially under the high-intensity photoexcitation. In such a case, the acquisition of the actual form of $n(\omega, z)$ becomes crucial, but its analytical determination is a challenging task. Also, in the multilayer analysis we assume that the photoexcitation affects the spectral properties only within the pump penetration depth of the sample. However, the heating simulation suggests that the effect of photoexcitation in terms of the rise in temperature affects much deeper into the sample than the penetration depth (see \textit{e.g.} Fig.~\ref{fig:figure_3}(a)). To examine the photoexcited state more accurately in bulk samples, the pump-probe penetration depth mismatch should be resolved carefully.
	
	Before concluding, it is worth mentioning again that the modeling of the sample heating in this study is not fully applicable in the timescale within the relaxation time where the system has not reached quasi-thermal equilibrium. Indeed, in our result at \SI{3}{\pico\second} (Fig.~\ref{fig:figure_2}(a)), we observed a difference between the experimental result and the simulation. In addition, since the specific heat of phonons becomes significantly larger with increasing temperature, the effect of pump-induced temperature increase should be infinitesimal in higher equilibrium temperature, including the measurements performed above $T_\mathrm{c}$, and cannot account for the emergence of JPR-like structure above $T_\mathrm{c}$ as observed in the stripe-phase \lbco{}~\cite{Nicoletti2014,Casandruc2015}.
	\section{Summary}
	We investigated the photoexcited nonequilibrium response of the high-$T_\mathrm{c}$ cuprate superconductor \lsco{}~$(x = 0.14)$ by utilizing near-infrared optical pump-THz probe spectroscopy. In the superconducting state, the continuous redshift of JPR was observed with increasing the photoexcitation densities, indicating the destruction of superconductivity. We demonstrated that the quasiequilibrium spectral feature appearing after photoexcitation is reproduced considering the sample surface heating induced by the pump energy. We also pointed out that conventional pump-probe analysis to extract the response function of photoexcited sample surface in optical pump-THz probe experiments can present serious artifacts in the transient spectra.
	
	\begin{acknowledgments}
	This work was supported in part by JSPS KAKENHI (Grants Nos. JP15H02102, JP26247057, and JP15H05452) and by Mitsubishi Foundation.
	\end{acknowledgments}
	\appendix
	\section{Details of the data acquision and analysis in the pump-probe measurement}\label{sec:pumpprobe}
	The pump-induced change of the photoexcited signal at the pump-probe delay time $t_\mathrm{pp}$ was obtained by utilizing the optical chopper to measure simultaneously the time-domain signal with the optical pump $E_\mathrm{wp}(t; t_\mathrm{pp})$ and without pump $E_\mathrm{np}(t; t_\mathrm{pp})$. By performing Fourier transform we obtained complex spectra $E_\mathrm{wp}(\omega; t_\mathrm{pp})$ and $E_\mathrm{np}(\omega; t_\mathrm{pp})$. We calculated the nonequilibrium complex reflectivity $r_\mathrm{pp}(\omega; t_\mathrm{pp})$ by multiplying the pump-induced change to the equilibrium complex reflectivity $r_\mathrm{eq}(\omega)$,
	\begin{align}
	r_\mathrm{pp}(\omega; t_\mathrm{pp}) = \frac{E_\mathrm{wp}(\omega; t_\mathrm{pp})}{E_\mathrm{np}(\omega;t_\mathrm{pp})}r_\mathrm{eq}(\omega). \label{eq:pp_refl}
	\end{align}
	Using $r_\mathrm{pp}(\omega; t_\mathrm{pp})$, we obtained the transient reflectivity $R_\mathrm{pp}(\omega; t_\mathrm{pp}) = |r_\mathrm{pp}(\omega; t_\mathrm{pp})|^2$, which are presented in Figs.~\ref{fig:figure_2} and \ref{fig:figure_4}(a).
	
	The key assumption in Eq.~(\ref{eq:pp_refl}) is that the spectral feature of $E_\mathrm{np}(\omega; t_\mathrm{pp})$ is identical to the spectrum measured in equilibrium state $E_\mathrm{eq}(\omega)$.
	However, if the modulation frequency of optical chopper for the optical pump is too fast under high pump fluence, the sample presents accumulative temperature increase because the next optical pump arrives at the sample before the sample temperature recovers back to its original equilibrium temperature. In such a case, the identity between $E_\mathrm{np}(\omega; t_\mathrm{pp})$ and $E_\mathrm{eq}(\omega)$ is violated. In order to avoid this average heating effect, we used two optical choppers with the modulation frequency of \SI{100}{\hertz} to limit the pump arrival to once in every \SI{10}{\milli\second}. 
	\section{Details of the heating simulation}\label{sec:heatsim}
	We discuss the procedure of simulation to account for the surface heating of the sample in detail. For this simulation, we consider that all of the energy injected by the intense optical excitation contributes to the sample heating in the quasiequilibrium state. The energy density of the optical pump $I(z)$ can be related with the excitation fluence by assuming an exponential decay by,
	\begin{align}
	I(z) &=(1-R)\qty(-\dv{F_\mathrm{pump}(z)}{z}) \notag\\
	&= (1-R)\frac{F_0}{d_\mathrm{pump}}e^{-z/d_\mathrm{pump}},
	\end{align} 
	where $F_0$ is the pump fluence, $d_\mathrm{pump} \approx \SI{660}{\nano\meter}$ is the penetration depth of the optical pump at \SI{800}{\nano\meter} calculated from the literature values~\cite{Uchida1996}, and $R\approx0.15$ is the reflectivity at \SI{800}{\nano\meter}~\cite{Uchida1996}. 
	The absorbed energy $I(z)$ and the temperature increase can be related by utilizing specific heat, which can be described as,
	\begin{align}
	C(T) = \gamma_0T + \alpha T^2 + \beta T^3 + \delta T^5,
	\end{align}
	where $\gamma_0T$ is the electronic specific heat, $\alpha T^2$ is the specific heat arising from the $d$-wave superconductivity, $\beta T^3 + \delta T^5$ is the lattice specific heat including the contributions up to $T^5$ term, respectively. We used the following  specific heat coefficients $\gamma_0= \SI{1.90}{\milli\joule\per\mole\per\kelvin\tothe{2}},\ \alpha=\SI{0.177}{\milli\joule\per\mole\per\kelvin\tothe{3}},\ \beta = \SI{0.120}{\milli\joule\per\mole\per\kelvin\tothe{4}}$ and $\delta= \SI{0.00093}{\milli\joule\per\mole\per\kelvin\tothe{6}}$, respectively, which are taken from the fitting result of the equilibrium specific measurement of the very near doping of \lsco{}~$(x = 0.15)$~\cite{Wen2004}.
	The quasiequilibrium temperature $T_\mathrm{samp}(z)$ can now be related to $I(z)$ through the following integral equation,
	\begin{align}
	I(z) = \int_{T_\mathrm{eq}}^{T_\mathrm{samp}(z)}C(T')\,\dd T',\label{eq:int_ItoT}
	\end{align}
	where $T_\mathrm{eq}$ is the equilibrium temperature. By numerically solving Eq.~(\ref{eq:int_ItoT}), we obtained $T_\mathrm{samp}(z)$ for various $F_0$. The construction of spatial distribution of complex refractive index $n(\omega, z)$ was then carried out by assigning equilibrium complex refractive index to each $z$ with corresponding temperature. The temperature-dependent equilibrium complex refractive index was obtained by a linear interpolation of the equilibrium result (Figs.~\ref{fig:figure_1}{(c)-(e)}) with respect to temperature.
	
	\section{Calculation of the effective refractive index $n_\mathrm{eff}(\omega)$}\label{sec:n_eff}
	\begin{figure}
		\includegraphics[width=\columnwidth]{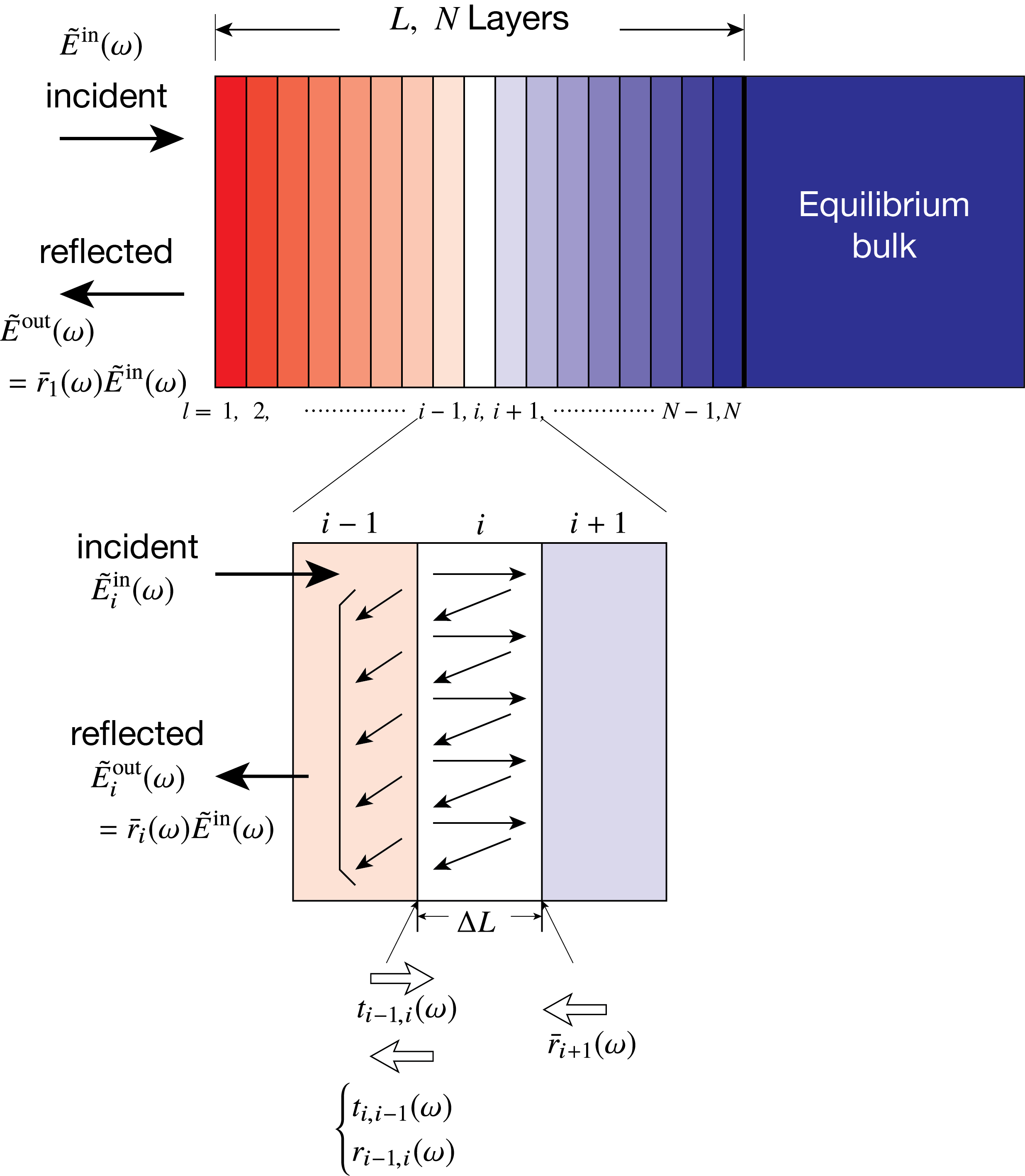}
		\caption{(Color online)~A schematic of the calculation procedure of the effective complex refractive index $n_\mathrm{eff}(\omega)$.}
		\label{fig:figure_effective}
	\end{figure}
	
	Here, we briefly discuss the method to calculate the effective complex refractive index $n_\mathrm{eff}(\omega)$ from the spatially non-uniform distribution of refractive index along the $z$-axis $n(\omega, z)$, which has been utilized both in the multilayer analysis and the heating simulation.
	In order to numerically solve $n_\mathrm{eff}(\omega)$, we consider the surface region of the sample as a stack of thin-film layers and calculate the reflectivity coefficients by considering infinite multiple reflections in each thin film, as schematically drawn in Fig.~\ref{fig:figure_effective}. In our current study, we split the sample surface of length $L=\SI{30}{\micro\meter}$ into $N = 1000$ layers. We denote the thickness of the each thin-film layer as $\Delta L (=L/N)$ and write $n_l(\omega) = n(\omega, l\Delta L)$ to indicate the complex refractive index of the $l$-th layer. 
	When we consider the reflection coefficient of the superconducting bulk to the deepest $N$-th layer, we use,
	\begin{align}
	\bar{r}_\mathrm{bulk}(\omega) = \frac{n_N(\omega) - n_\mathrm{eq}(\omega)}{n_N(\omega) + n_\mathrm{eq}(\omega)},
	\end{align}
	where $n_\mathrm{eq}(\omega)$ is the equilibrium complex refractive index.
	Now, when we include the $N$-th layer, the reflected THz probe acquires the following reflection coefficient,
	\begin{align}
	\bar{r}_{N}(\omega) &= r_{N-1, N}(\omega)\notag\\
	&\qquad+ t_{N-1, N}(\omega) \bar{r}_\mathrm{bulk}(\omega)t_{N, N-1}(\omega)e^{i\Phi_N(\omega)}\notag\\
	&\qquad\qquad\times\sum_{k=0}^\infty\left[r_{N,N-1}(\omega)\bar{r}_\mathrm{bulk}(\omega)e^{i\Phi_N(\omega)}\right]^k \notag\\
	&=\frac{r_{N-1, N}(\omega) + \bar{r}_\mathrm{bulk}(\omega)e^{i\Phi_N(\omega)}}{1 + r_{N-1, N}(\omega)\bar{r}_\mathrm{bulk}(\omega)e^{i\Phi_N(\omega)}},
	\end{align}
	where, 
	\begin{align}
	t_{i,j}(\omega) &= \frac{2n_i(\omega)}{n_i(\omega) + n_j(\omega)}, \\
	r_{i,j}(\omega)&= \frac{n_i(\omega) - n_j(\omega)}{n_i(\omega) + n_j(\omega)}
	\end{align}
	are the Fresnel coefficients for the transmittance and reflection at the boundary of two layers with $n_i(\omega)$ and $n_j(\omega)$, and $\Phi_i(\omega) = 2n_i(\omega)\omega\Delta L/c$ gives the phase factor of the multiple reflection.
	Similarly, the effective Fresnel's reflection coefficient at the $l$-th layer $\bar{r}_l(\omega)$ can be calculated using the $\bar{r}_{l+1}(\omega)$,
	\begin{align}
	\tilde{r}_l(\omega) = \frac{r_{l-1, l}(\omega) + \bar{r}_{l+1}(\omega)e^{i\Phi_l(\omega)}}{1 + r_{l-1, l}(\omega)\bar{r}_{l+1}(\omega)e^{i\Phi_l(\omega)}}.
	\end{align}
	By repeating this calculation until we reach the surface layer $l =1$, we obtained the total effective Fresnel coefficient $\bar{r}_1(\omega)$. For the calculation of $\bar{r}_1(\omega)$, we adopt the refractive index of the sample atmosphere to be $n_\mathrm{air}(\omega) = n_0(\omega) = 1$. Finally, we can calculate $n_\mathrm{eff}(\omega)$ as
	\begin{align}
	n_\mathrm{eff}(\omega) = \frac{1 - \bar{r}_1(\omega)}{1 + \bar{r}_1(\omega)}.
	\end{align}
	
	\bibliographystyle{apsrev4-1}
	\bibliography{manuscript_bib}
	
\end{document}